\documentclass[aip, pof, amsmath, amssymb, reprint]{revtex4-1}

\usepackage{color}
\usepackage{amssymb}
\usepackage{graphicx}
\usepackage{amsmath}
\usepackage{subfig}
\usepackage{booktabs}
\usepackage{tikz}
\usepackage{pgfplots}
\usepgfplotslibrary{external}
\tikzset{external/system call={lualatex \tikzexternalcheckshellescape -halt-on-error -interaction=batchmode -jobname "\image" "\texsource"}}
\usepackage{placeins}
\usepackage{morefloats}

\def\XXint#1#2#3{{\setbox0=\hbox{$#1{#2#3}{\int}$ }
\vcenter{\hbox{$#2#3$ }}\kern-.6\wd0}}

\begin{document}
\title{Wake angle for surface gravity waves on a finite depth fluid}

\author{Ravindra Pethiyagoda}
\author{Scott W. McCue}
 \email{scott.mccue@qut.edu.au}
\author{Timothy J. Moroney}
\affiliation{Mathematical Sciences, Queensland University of Technology, Brisbane QLD 4001, Australia}
\date{\today}
\begin{abstract}
Linear water wave theory suggests that wave patterns caused by a steadily moving disturbance are contained within a wedge whose half-angle depends on the depth-based Froude number $F_H$.  For the problem of flow past an axisymmetric pressure distribution in a finite-depth channel, we report on the apparent angle of the wake, which is the angle of maximum peaks.  For moderately deep channels, the dependence of the apparent wake angle on the Froude number is very different to the wedge angle, and varies smoothly as $F_H$ passes through the critical value $F_H=1$.  For shallow water, the two angles tend to follow each other more closely, which leads to very large apparent wake angles for certain regimes.
\end{abstract}
\maketitle

The linear dispersion relation for small amplitude water waves gives rise to intricate and distinctive surface wave patterns that have been the subject of enduring interest to applied mathematicians and physicists since the 19th century\cite{Craik2004}.  When these patterns form at the wake of a steadily moving object, they are often referred to as ship wave patterns or ship wakes, regardless of whether or not the object is an actual ship.  Relatively simple geometric arguments\cite{Lighthill1978} show that ship wave patterns propagate within a travelling wedge; for an infinitely deep fluid, this wedge has a half-angle of $\theta_\mathrm{wedge}=\arcsin(1/3)\approx 19.47^\circ$ (referred to here as the Kelvin angle) regardless of the speed of the disturbance\cite{kelvin87}.

For a fluid of finite depth, the derivation of the wedge angle is more complicated\cite{havelock08}, and turns out to depend on the speed of the disturbance $U$, which itself is measured by the depth-based Froude number $F_H=U/\sqrt{gH}$.  Here $H$ is the depth of the fluid and $g$ is acceleration due to gravity.  Of particular interest here, the wedge angle has the following properties: $\theta_\mathrm{wedge}\rightarrow \arcsin(1/3)^+$ as $F_H\rightarrow 0^+$; $\theta_\mathrm{wedge}\rightarrow \pi/2^-$ as $F_H\rightarrow 1^-$; $\theta_\mathrm{wedge}\rightarrow \pi/2^-$ as $F_H\rightarrow 1^+$; $\theta_\mathrm{wedge}\sim F_H^{-1}$ as $F_H\rightarrow\infty$.  We note that for $F_H<1$, the flow is referred to as being subcritical.  Subcritical flows are characterised by two types of waves, transverse and divergent.  Transverse waves run roughly perpendicular to the direction the disturbance is travelling.  Divergent waves propagate at an acute angle to this direction.  For supercritical flows, $F_H>1$, the wave pattern is entirely made up of divergent waves.

There has been a resurgence of interest in ship wave patterns\cite{torsvik15}, with a particular focus on the wake angle that is actually observed in practice\cite{Dias2014}.  Here we pay attention to what we call the apparent wake angle, $\theta_\mathrm{app}$, which is the angle between the line passing through the highest peaks of the wave and the centreline.  For axially symmetric pressure distributions acting on the free surface of an infinitely deep fluid, numerical simulations show that the apparent wake angle is slightly less than the Kelvin angle $\arcsin(1/3)$ for moderately fast disturbances; however, for sufficiently fast-moving disturbances, the apparent wake angle appears to decrease like the inverse of the length-based Froude number, $F_L=U/\sqrt{gL}$, where $L$ is a length-scale associated with the pressure distribution\cite{rabaud13,darmon14} (see also Refs~\cite{barnell86,carusotto13}).  Further recent studies have been undertaken on the effect of non-axisymmetric pressure distributions\cite{benzaquen14,moisy14,noblesse14,rabaud14}, a continuous distribution of sources\cite{Zhang15}, constant vorticity\cite{ellingsen14}, nonlinearity\cite{pethiyagoda14a,pethiyagoda14b} and surface tension\cite{moisy14b}. A comparison of the results in Refs~\cite{rabaud13,darmon14,noblesse14} is given by He \emph{et al.}~\cite{he14}. All of these studies are for an infinite-depth fluid.

By considering the problem of flow past an axially symmetric pressure distributions in a fluid of {\em finite} depth, we shed light on the seemingly contradictory results for the wedge angle and recently reported apparent wake angle.  As noted above, the wedge angle $\theta_\mathrm{wedge}$ increases from $\arcsin(1/3)$ to $\pi/2$ as $F_H$ increases to $F_H=1$, and then monotonically decreases for $F_H>1$.  However, for a moderately deep fluid, we find the dependence of the apparent angle $\theta_\mathrm{app}$ on $F_H$ is very different, and in fact behaves much like the infinite depth case.  This is remarkable because the transition between subcritical ($F_H<1$) and supercritical ($F_H>1$) flows is characterised by distinct changes in qualitative behaviour in the wave pattern and the singularity structure of the dispersion relationship, while the apparent wake angle $\theta_\mathrm{app}$ changes smoothly as $F_H$ increases through $F_H=1$.
On the other hand, for shallow water, the apparent angle $\theta_\mathrm{app}$ follows the wedge angle $\theta_\mathrm{wedge}$ closely, giving rise to very large apparent wake angles for moderately small length-based Froude numbers, $F_L$.

In the following, we consider an axisymmetric pressure distribution of strength $P_0$ and intrinsic length $L$ acting on the surface of a fluid flowing in the positive $x$ direction with speed $U$. We will present solutions for a fluid of infinite depth and a finite depth $H$.  To enable a direct comparison with recent work\cite{rabaud13,darmon14}, our dimensional pressure distribution is $p(x,y)=P_0\,\mathrm{exp}(-\pi^2(x^2+y^2)/L^2)$.  For the infinite depth problem we choose to scale speeds by $U$ and lengths by $L$.  This leads to two nondimensional parameters, $\epsilon_L=P_0/(\rho g L)$, a measure of pressure strength (where $\rho$ is the fluid density), and the length-based Froude number, $F_L$.  Using linear theory, an exact solution for the free-surface $\zeta(x,y)$ in the infinite depth case is\cite{wehausen60}:
\begin{align}
\zeta(x,y) = &-\epsilon_L p(x,y)+\frac{\epsilon_L}{2\pi^2} \int\limits_{-\pi/2}^{\pi/2}\,\int\limits_{0}^{\infty}\frac{k^2\tilde{p}(k,\psi)\cos(k[|x|\cos\psi+y\sin\psi])}{k-k_0}\,\,\mathrm{d}k\,\,\mathrm{d}\psi\notag\\
&-\frac{\epsilon_L H(x)}{\pi} \int\limits_{-\pi/2}^{\pi/2}k_0^2\tilde{p}(k_0,\psi)\sin(k_0[x\cos\psi+y\sin\psi])\,\,\mathrm{d}\psi,\label{eq:exactLinearInfinite}
\end{align}
where $\tilde{p}(k,\psi)=\delta^2 \mathrm{exp}(-\delta^2k^2/(4\pi^2))/\pi$ is the Fourier transform of our pressure distribution and $H(x)$ is the Heaviside function.  The path of $k$-integration is diverted below the pole $k=k_0$, where $k_0=1/(F_L^2\cos^2\psi)$.

For the finite depth problem we instead use $H$ as the length scale, which leads to three nondimensional parameters: the pressure strength $\epsilon_H=P_0/(\rho g H)$, the depth-based Froude number $F_H$, and the ratio of pressure length scale to fluid depth $\delta=L/H$.  The nondimensional parameters for the finite and infinite depth problems are related by $F_L=F_H/\sqrt{\delta}$ and $\epsilon_L=\epsilon_H/\delta$.   The exact solution for the finite-depth problem is\cite{wehausen60}
\begin{align}
\zeta(x,y) = &-\epsilon_H p(x,y)+\frac{\epsilon_H F_H^2}{2\pi^2}\int\limits_{-\pi/2}^{\pi/2}
\,\,\int\limits_{0}^{\infty}\frac{k^2\tilde{p}(k,\psi)\cos(k[|x|\cos\psi+y\sin\psi])}{kF_H^2-\sec^2\psi\tanh k}\,\,\mathrm{d}k\,\mathrm{d}\psi\notag\\
&-\frac{2\epsilon_H F_H^2 H(x)}{\pi}\int\limits_{\psi_0}^{\pi/2}\frac{k_1^2\tilde{p}(k_1,\psi) \sin(k_1x\cos\psi)\cos(k_1y\sin\psi)}{F_H^2-\sec^2\psi\,\mathrm{sech}^2k_1}\,\mathrm{d}\psi,\label{eq:exactLinearFinite}
\end{align}
where $\psi_0=0$ for $F_H<1$ and $\psi_0=\arccos(1/F_H)$ for $F_H>1$.  As with the infinite depth solution (\ref{eq:exactLinearInfinite}) the path of $k$-integration is diverted below the pole, which this time is denoted by $k=k_1(\psi)$, where $k_1$ is the real positive root of
\begin{equation}
kF_H^2-\sec^2\psi\tanh k=0,\qquad \psi_0<\psi<\frac{\pi}{2}.
\label{dispersion}
\end{equation}
In both the infinite-depth (\ref{eq:exactLinearInfinite}) and finite-depth (\ref{eq:exactLinearFinite}) solutions, the double integral rapidly tends to zero far away from the pressure distribution. Therefore, to analyse the wake in the far field, only the single integral term is required.

To measure the apparent wake angle, $\theta_\mathrm{app}$, we need to locate the line that passes through the highest peaks of the wave, and then calculate the angle between that line and the centreline $y=0$.  The method for locating the highest peaks depends on the flow regime. For subcritical flows (and the infinite depth case), we isolate each transverse wave, mark the highest point, and fit a line through the highest points of all the wavelengths (we call this method 1, used previously in Refs~\cite{darmon14,pethiyagoda14b}, for example).  For supercritical flows, there are no transverse waves, so we instead use the highest points on each of the divergent waves to fit the line (method 2).

\begin{figure}[htb]
\centering
\includegraphics[width=0.98\linewidth]{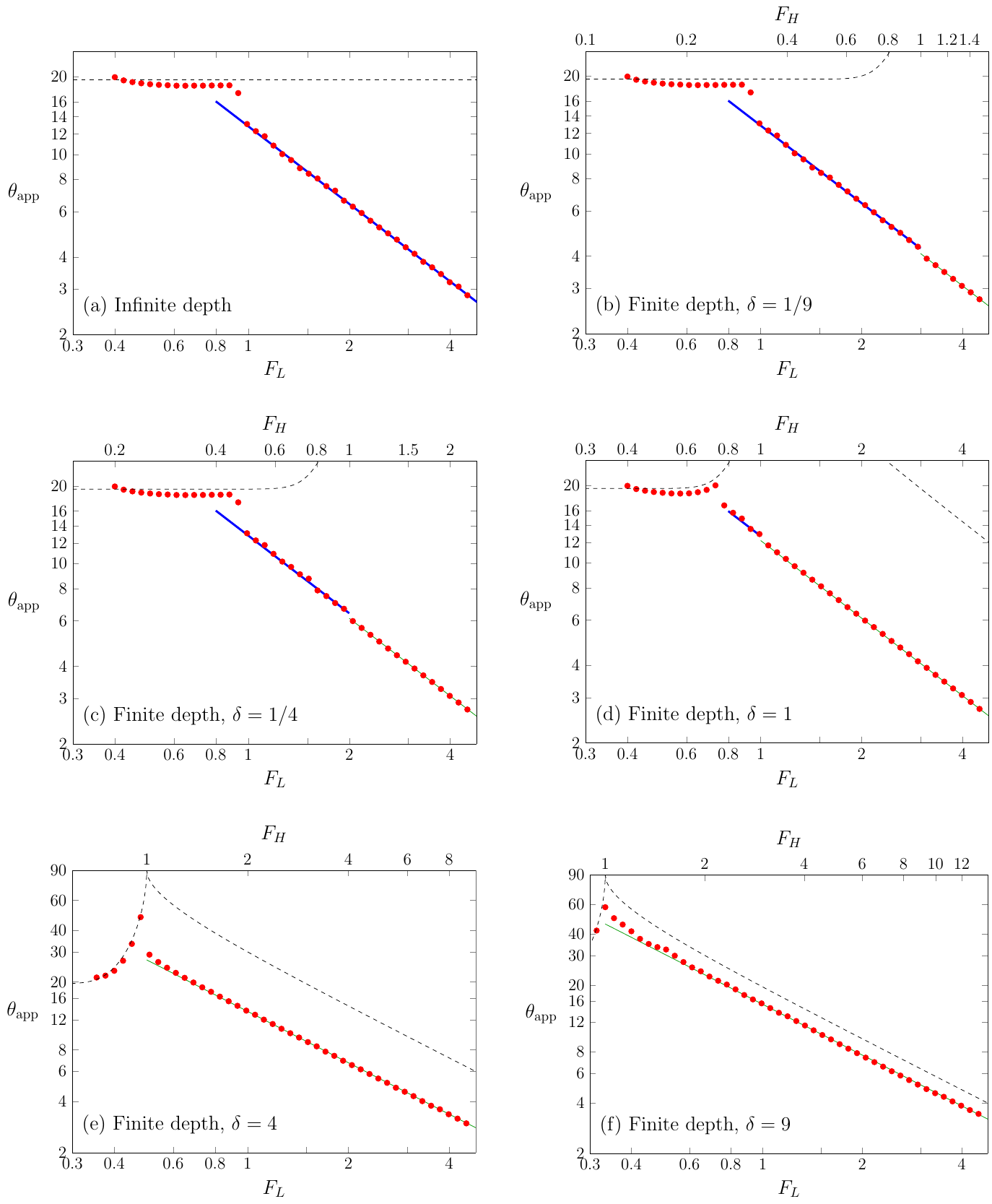}
\caption{Measured apparent wake angles $\theta_\mathrm{app}$ in degrees (solid circles) plotted against the Froude number for the linear solutions (\ref{eq:exactLinearInfinite}) and (\ref{eq:exactLinearFinite}) for a fluid of (a) infinite depth and (b)--(f) finite depth, presented on a log-log scale. The dashed line is (a) the Kelvin angle $\arcsin(1/3)\approx 19.47^\circ$ and (b)--(f) the wedge angle $\theta_\mathrm{wedge}$ in the finite-depth case.  The thick (blue) line is the theoretical asymptote (\ref{eq:asymFunc}) given by method 1 (valid for subcritical flows) and the thin (green) line is asymptote given by method 2 (required for supercritical flows).}
\label{fig:WakeVsFrL}
\end{figure}

In Fig.~\ref{fig:WakeVsFrL} we present data for apparent wake angle against both Froude numbers $F_L$ and $F_H$ for a variety of dimensionless pressure length scales $\delta$.  These data are represented by the (red) solid dots.  As just mentioned, the data for the infinite depth case and for the subcritical flows $F_H<1$ are calculated with method 1, while data for supercritical flows $F_H>1$ are calculated with method 2.  We see in Fig.~\ref{fig:WakeVsFrL}(b)-(d) that for moderately deep channels, the qualitative behaviour is very similar to the infinite depth case in Fig.~\ref{fig:WakeVsFrL}(a) considered previously\cite{darmon14}; that is, the apparent wake angle $\theta_\mathrm{app}$ is approximately $\arcsin(1/3)$ for $F_L$ less than some value (roughly $F_L=1$ or slightly less), then decreases like $F_L^{-1}$ for large values of $F_L$.  Also plotted in each panel of Fig.~\ref{fig:WakeVsFrL} is $\theta_\mathrm{wedge}$ (a dashed line).  It is interesting to see that for deep to moderately deep channels, the apparent wake angle $\theta_\mathrm{app}$ does not follow $\theta_\mathrm{wedge}$.

\begin{figure}[htb]
\centering
\includegraphics[width=0.98\linewidth]{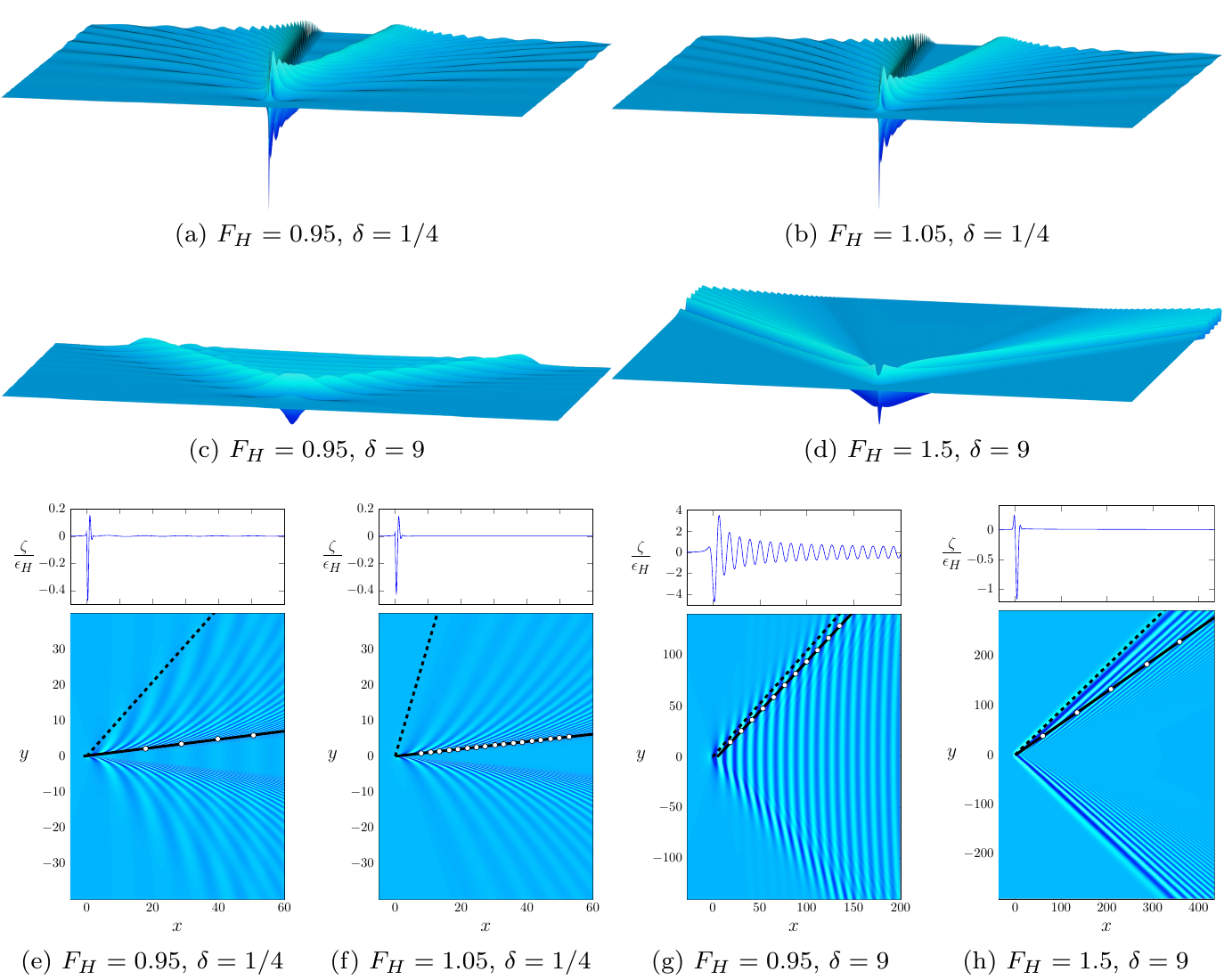}
\caption{(a)--(d) Free surface profiles for various values of $F_H$ and $\delta$. (e)--(h) A plan view of the profiles in (a)--(d) with a centreline plot. The open circles are the highest peaks and the solid line is the line of best fit. The dashed lines the wedge angle $\theta_\mathrm{wedge}$.}
\label{fig:freesurfaceProfiles}
\end{figure}

It is insightful to observe the wave patterns themselves in Fig.~\ref{fig:freesurfaceProfiles}, especially as the flow transitions from subcritical $F_H<1$ to supercritical $F_H>1$.  For sufficiently deep channels ($\delta$ sufficiently small), the transition has almost no effect on $\theta_\mathrm{app}$ (also refer back to Fig.~\ref{fig:WakeVsFrL}(b)-(d), where the slight kink in the data for $\theta_\mathrm{app}$ at $F_H=1$ is due to the change in method of measuring the angle).  In Fig.~\ref{fig:freesurfaceProfiles}(a)-(b), we present wave patterns for a typical small value (taken to be $\delta=1/4$), computed for $F_H$ slightly less than and slightly greater than the critical value $F_H=1$.  The corresponding plan view and centreline plots are shown in panels (e)-(f).  In the plan view, the highest peaks are represented by open circles, while the angle $\theta_\mathrm{wedge}$ is represented by the dashed lines.  These surface profiles help to show how, for sufficiently deep channels, the apparent wake angle $\theta_\mathrm{app}$ varies smoothly through $F_H=1$ and is not at all close to $\theta_\mathrm{wedge}$.

Returning to Fig.~\ref{fig:WakeVsFrL}, for shallow channels, as in Fig.~\ref{fig:WakeVsFrL}(e)-(f), there are clear differences between supercritical and subcritical flows.  In the former regime ($F_H>1$, $\delta\gg 1$), the trend in the data is dominated by the $F_L^{-1}$ decay.  While for the latter regime ($F_H<1$, $\delta\gg 1$), the apparent wake angle $\theta_\mathrm{app}$ can be very large (much larger than $\arcsin(1/3)$), and here $\theta_\mathrm{app}$ {\em does} follow $\theta_\mathrm{wedge}$ closely.  The transition between subcritical and supercritical flows has a dramatic qualitative effect on the free surface (Fig.~\ref{fig:freesurfaceProfiles}(c),(d)(g),(h)) with the absence of the transverse waves for $F_H>1$, exemplified in the centreline plots of Fig.~\ref{fig:freesurfaceProfiles}(g),(h).

To calculate the asymptotes for the wake angle $\theta_\mathrm{app}$ in Fig.~\ref{fig:WakeVsFrL} (the thick blue line and the thin green line) we perform a far-field stationary phase approximation on the second integral in (\ref{eq:exactLinearFinite}) using polar coordinates $(r,\theta)$ (defined via $x=r\cos\theta$, $y=r\sin\theta$) to obtain
\[
\zeta(r,\theta)\approx a(r,\theta)\sin(rg(\psi_1(\theta),\theta)-\pi/4),
\]
where $g(\psi,\theta)=k_1(\psi)\cos(\psi+\theta)$ and
\begin{align}
a(r,\theta)=\frac{\mathrm{const}}{r^{1/2}}\,\frac{k_1(\psi_1(\theta))^2\exp\left(-\delta^2k_1(\psi_1(\theta))^2/(4\pi^2)\right)}{(F_H^2-\sec(\psi_1(\theta))^2\mathrm{sech}(k_1(\psi_1(\theta)))^2)
\sqrt{g_{\psi\psi}(\psi_1(\theta),\theta)}}.
\label{eq:sphasebits}
\end{align}
Recall $k_1$ is the real positive root of (\ref{dispersion}).  The function $\psi_1(\theta)$ satisfies
\[
k_1'(\psi_1)\cos(\psi_1+\theta)-k_1(\psi_1)\sin(\psi_1+\theta)=0,\qquad \psi_0<\psi_1<\frac{\pi}{2}\,\,\mathrm{and}\,\,0<\theta<\theta_\mathrm{wedge},
\]
while $g_{\psi\psi}$ is the second partial derivative of $g$ with respect to $\psi$.

We must now find the maximum of $a(r,\theta)$.  The two approaches depend on the method used to measure the wake angle $\theta_\mathrm{app}$.  For method 1, $r$ is held constant (to $r=1$, say) and we then compute the turning point of $a(1,\theta)$. For method 2 we must choose $r$ such that $\sin(rg(\psi_1(\theta),\theta)-\pi/4)=1$, which leads to the new amplitude function
\begin{equation}
a(g(\psi_1,\theta)^{-1},\psi)=\frac{\mathrm{const}\,k_1(\psi_1(\theta))^2\exp\left(-\delta^2k_1(\psi_1(\theta))^2/(4\pi^2)\right)}{(F_H^2-\sec(\psi_1(\theta))^2\mathrm{sech}(k_1(\psi_1(\theta)))^2}\sqrt{\frac{g(\psi_1(\theta),\theta)}
{g_{\psi\psi}(\psi_1(\theta),\theta)}}.\label{eq:amplitude}
\end{equation}

\begin{figure}[htb]
\centering\includegraphics[width=.9\linewidth] {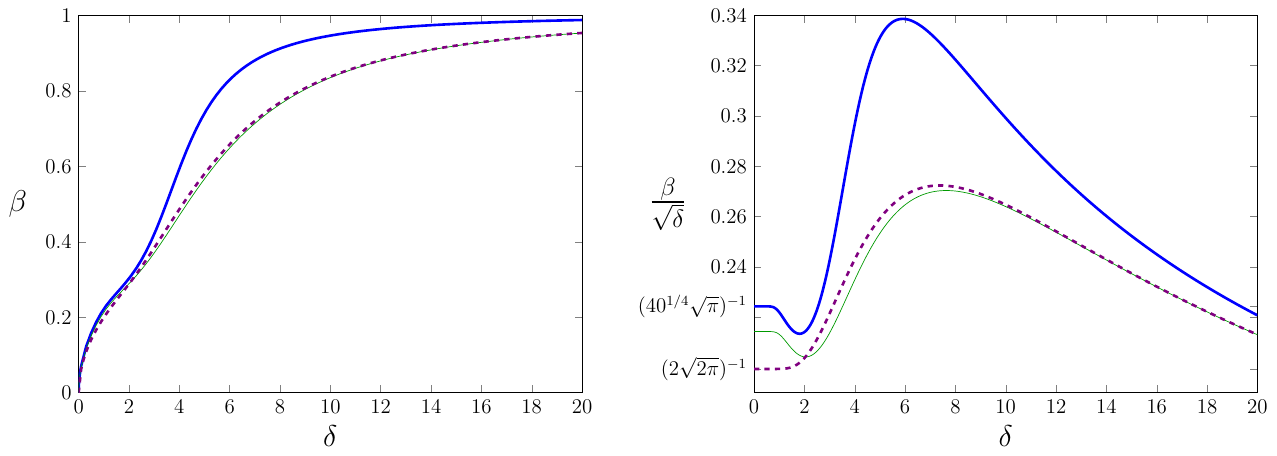}
\caption{Plots of (a) $\beta$ and (b) $\beta/\sqrt{\delta}$ in (\ref{eq:asymFunc}) against $\delta$. The thick (blue) line comes from method 1, the thin (green) line from method 2, the dashed (purple) line by (\ref{eq:kappa}).}
\label{fig:kappaVdelta}
\end{figure}

A simple numerical test shows that, for each value of $\delta$, the maximum of both (\ref{eq:sphasebits}) and (\ref{eq:amplitude}) behaves like $F_H{\theta}_\mathrm{app}\rightarrow \,\mathrm{const}$ as $F_H\rightarrow\infty$, so that, in both cases,
\begin{equation}
\theta_\mathrm{app}\sim\frac{\beta}{F_H}=\frac{\beta}{\sqrt{\delta}F_L}
\quad\mbox{as}\quad F_H\rightarrow\infty.
\label{eq:asymFunc}
\end{equation}
The numerical values of $\beta$ for both methods are found to increase from $\beta=0$ at $\delta=0$ to $\beta\rightarrow 1^-$ as $\delta\rightarrow\infty$, thus the asymptotic behaviour of the wake angle is dominated by finite depth effects and approaches the finite depth wedge angle $\theta_\mathrm{wedge}$. As shown in Fig.~\ref{fig:kappaVdelta}, we note that method 1 gives $\beta/\sqrt{\delta}\rightarrow 1/(40^{1/4}\sqrt{\pi})$ as $\delta\rightarrow 0$, which agrees with the infinite depth result computed by Darmon et al.~\cite{darmon14}.

These results are worth comparing with the phenomenological model of Moisy and Rabaud\cite{moisy14,moisy14b}, which for the present finite-depth problem can be extended in a straightforward fashion to predict that
\begin{equation}
\tan\theta_\mathrm{app}=c_g(1-c_p^2)^{1/2}/(1-c_gc_p)\quad\mbox{with}\quad k=2\pi/\delta,
\label{eq:moisy}
\end{equation}
where $c_g$ and $c_p$ are the group and phase velocities, respectively, defined by $c_g=\mathrm{d}\omega/\mathrm{d}k$ and $c_p=\omega/k$, and the dispersion relation is $\omega=(k\tanh k)^{1/2}/F_H$.  As $c_g=O(F_H^{-1})$ and $c_p=O(F_H^{-1})$ as $F_H\rightarrow\infty$, we see that Moisy and Rabaud's method gives $\theta_\mathrm{app}\sim c_g$ with $k=2\pi/\delta$ which, using the notation in (\ref{eq:asymFunc}), gives
\begin{equation}
\beta=\frac{\delta\tanh(2\pi/\delta)+2\pi\mathrm{sech}^2(2\pi/\delta)}{2\sqrt{2\pi\delta\tanh(2\pi/\delta)}}. \label{eq:kappa}
\end{equation}
This result is also shown in Fig.~\ref{fig:kappaVdelta} (dashed line).  In the deep water limit, (\ref{eq:kappa}) gives $\beta/\sqrt{\delta}\rightarrow(2\sqrt{2\pi})^{-1}$ as $\delta\rightarrow 0$, which agrees with Ref\cite{moisy14}.  For shallow water $\delta\gg 1$, it collapses to our method 2.

\begin{figure}[htb]
\centering
\includegraphics[width=.90\linewidth]{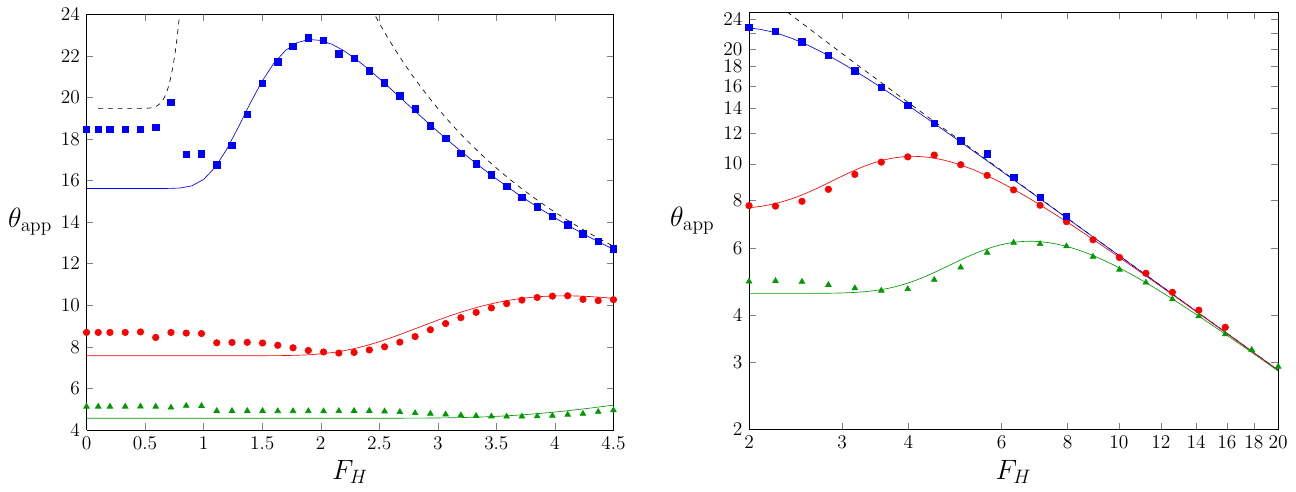}
\caption{Measured apparent wake angles $\theta_\mathrm{app}$ in degrees plotted against the depth-based Froude number $F_H$ for the linear solution (\ref{eq:exactLinearFinite}) for constant values of $F_L=0.7$ (blue squares), $1.5$ (red dots) and  $2.5$ (green triangles). Note that as $F_H$ increases, $\delta$ increases such that $F_H/\sqrt{\delta}=F_L$. The solid lines represent (\ref{eq:moisy}), while the dashed line indicates the wedge angle $\theta_\mathrm{wedge}$.}
\label{fig:LinConstFrL}
\end{figure}

We now plot in Fig.~\ref{fig:LinConstFrL} the apparent wake angle $\theta_\mathrm{app}$ against $F_H$ for fixed values of $F_L$.  Note that for each fixed value of $F_L$, an increase in $F_H$ must be associated with an increase in $\delta$ so that $F_L=F_H/\sqrt{\delta}$.  If we imagine a given pressure distribution with dimensional length scale $L$ moving at dimensional speed $U$, then each set of data points in Fig.~\ref{fig:LinConstFrL} corresponds to decreasing the channel depth from infinite ($F_H=0$) to zero ($F_H=\infty$).  With this in mind, Figure \ref{fig:LinConstFrL}(a) shows that for a faster moving stream (see data for $F_L=1.5$ and $2.5$), the transition from subcritical to supercritical flow via a decreasing fluid depth has little effect on the apparent wake angle $\theta_\mathrm{app}$. Ultimately, regardless of the speed of the stream (that is, for all fixed values of $F_L$), the apparent wake angle asymptotes to the wedge angle $\theta_\mathrm{wedge}$ for sufficiently shallow depths (see Figure \ref{fig:LinConstFrL}(b) for large $F_H$).  It is interesting to compare these results with the phenomenological model (\ref{eq:moisy}); the agreement is not so good for deep channels, but improves as the depth decreases.

In summary, we have considered the prototype problem of flow past an axisymmetric pressure distribution acting on the surface of a finite-depth channel.  For the linearised solution, we have explored the dependence of the apparent wake angle $\theta_\mathrm{app}$ (or angle of maximum peaks) on the dimensionless variables.  This work extends previous studies for flow past an axisymmetric pressure distribution for the infinite depth case.  It has been insightful to fix the depth of the channel and increase the speed of the flow (fixed $\delta$, increasing $F_H$, see Figure \ref{fig:WakeVsFrL}) and also to fix the speed of the flow and decrease the depth (fixed $F_L$, increasing $F_H$, see Figure \ref{fig:LinConstFrL}).  It is remarkable that for much of the parameter space, the apparent wake angle $\theta_\mathrm{app}$ varies smoothly as $F_H$ is increased through critical value $F_H=1$.  On other hand, for shallow water, we see the apparent wake angle $\theta_\mathrm{app}$ tends to follow the wedge angle $\theta_\mathrm{wedge}$ closely, which for near critical flows produces very large apparent angles, much higher than the Kelvin angle $\arcsin(1/3)$.

We close by noting that the we have deliberately used a very simple pressure distribution, which more easily facilitates a comparison with many recent studies, but may not accurately reflect the wave patterns produced by a real ship.  An alternative simplified approach, for example, is to employ a pair of point pressures on the surface\cite{noblesse14} (one positive and one negative), or a continuous distribution of sources\cite{Zhang15}.  Indeed, the wave pattern generated by a finite-depth flow past a pair of point pressures was considered very recently by Zhu et al.~\cite{zhu15}.  This study found that for subcritical flows in very shallow channels, the wake angle was close to the wedge angle $\theta_\mathrm{wedge}$, which is very similar to our observations.  On the other hand, their argument based on interference effects suggests that the wake angle scales like $F_H^{-2}$ in the large Froude number limit, which is very different to (\ref{eq:asymFunc}).  This difference in the large Froude number scaling is, at least in part, likely due to the difference in the flow configurations considered (see the relevant discussions in Refs\cite{moisy14,he14} about the effect of non-axisymmetric pressure disturbances on the large Froude number scaling).  Either way, a further advantage of our approach here is that we provide direct comparisons between analytical predictions and measured values.

\vspace{2ex}
S.W.M. acknowledges the support of the Australian Research Council (DP140100933).  The authors thank one of the anonymous referees for pointing out Ref.~\cite{zhu15}.

\end{document}